\documentclass[twocolumn,nofootinbib]{revtex4-1}
\usepackage{amsmath,amssymb,bm,graphicx,hyperref}
\allowdisplaybreaks[1]

\renewcommand\d{\partial}
\newcommand\+{\dagger}

\renewcommand\>{\rangle}
\newcommand\up{\uparrow}
\newcommand\down{\downarrow}
\newcommand\0{\bm{0}}
\newcommand\e{\hat{\bm{e}}}
\newcommand\p{\bm{p}}
\renewcommand\r{\bm{r}}
\newcommand\x{\bm{x}}
\renewcommand\H{\mathcal{H}}

\begin{document}

\title{Quantum droplet of one-dimensional bosons with a three-body attraction}

\author{Yuta Sekino}
\author{Yusuke Nishida}
\affiliation{Department of Physics, Tokyo Institute of Technology,
Ookayama, Meguro, Tokyo 152-8551, Japan}

\date{November 2017}

\begin{abstract}
Ultracold atoms offer valuable opportunities where interparticle interactions can be controlled at will.
In particular, by extinguishing the two-body interaction, one can realize unique systems governed by the three-body interaction, which is otherwise hidden behind the two-body interaction.
Here we study one-dimensional bosons with a weak three-body attraction and show that they form few-body bound states as well as a many-body droplet stabilized by the quantum mechanical effect.
Their binding energies relative to that of three bosons are all universal and the ground-state energy of the dilute droplet is found to grow exponentially as $E_N/E_3\to\exp(8N^2/\sqrt3\pi)$ with increasing particle number $N\gg1$.
The realization of our system with coupled two-component bosons in an optical lattice is also discussed.
\end{abstract}

\maketitle

\section{Introduction}
Matter surrounding us is hierarchically structured.
Quarks are bound into nucleons, nucleons into nuclei, nuclei and electrons into atoms, atoms into molecules, and eventually matter is formed.
Understanding of how these constituents attract each other to form their complex in each hierarchy has been the central objective of physics.
While atomic and nuclear interactions induced by electric and pionic fields are usually approximated to be pairwise, interactions among three and more constituents intrinsically arise~\cite{Primakoff:1939,Axilrod:1943,Muto:1943,Fujita:1957}.
However, three- and higher-body interactions are generally weak and thus can often be treated as small perturbations in binding energies dominated by two-body interactions~\cite{Hammer:2013}.

From this perspective, ultracold atoms offer valuable opportunities where interparticle interactions can be controlled at will~\cite{Chin:2010}.
While interesting physics is usually aimed at by amplifying the two-body interaction, one can also extinguish it to realize unique systems governed by the three-body interaction~\cite{Daley:2014,Petrov:2014a,Petrov:2014b,Paul:2016}.
Whether such novel systems without two-body but with tunable three-body interactions exhibit interesting physics remains largely unexplored and thus can be an important research frontier in ultracold atoms.
One of the recent advances toward this direction was the discovery of the semisuper Efimov effect, where two-dimensional bosons at a three-body resonance form an infinite tower of four-body bound states with the universal scaling law in their binding energies~\cite{Nishida:2017}.

In this paper, we turn to one-dimensional bosons without two-body but with tunable three-body interactions.
We start by discussing the realization of our system with coupled two-component bosons in an optical lattice (Sec.~\ref{sec:lattice}) and then show that a weak three-body attraction leads to one three-body and three four-body bound states, whose binding energy ratios are universal (Sec.~\ref{sec:few-body}).
Furthermore, $N\gg1$ bosons form a dilute droplet stabilized by the quantum mechanical effect and its ground-state energy is found to grow exponentially with increasing particle number (Sec.~\ref{sec:many-body}).

\section{\label{sec:lattice}Lattice realization}
In order to realize one-dimensional bosons without two-body but with tunable three-body interactions, we employ the scheme proposed in Ref.~\cite{Petrov:2014b}.
Let us consider coupled two-component bosons in an optical lattice, which in the tight-binding approximation are described by the Hamiltonian
\begin{align}\label{eq:original}
H = -t_x\sum_{\sigma=\up,\down}\sum_{\<i,j\>}b_{\sigma i}^\+b_{\sigma j} + \sum_i\H_i,
\end{align}
with
\begin{align}\label{eq:on-site}
\H_i &= -\frac\Omega2(b_{\up i}^\+b_{\down i} + b_{\down i}^\+b_{\up i})
- \frac\Delta2(b_{\up i}^\+b_{\up i} - b_{\down i}^\+b_{\down i}) \notag\\
&\quad + \sum_{\sigma,\sigma'}\frac{g_{\sigma\sigma'}}{2}
b_{\sigma i}^\+b_{\sigma'i}^\+b_{\sigma'i}b_{\sigma i},
\end{align}
where $t_x$ is the intersite tunneling amplitude, $\Omega$ the Rabi frequency, $\Delta$ the detuning, $g_{\sigma\sigma'}$ the on-site interaction energy, and $\hbar=1$.
As far as low-energy physics relative to the spin gap $\sqrt{\Omega^2+\Delta^2}$ is concerned, the higher-energy spin component can be integrated out to reduce the original problem in Eq.~(\ref{eq:original}) to an effective single-component problem described by
\begin{align}\label{eq:effective}
H_\mathrm{eff} = -t_x\sum_{\<i,j\>}b_i^\+b_j
+ \sum_i\sum_{n=1}^\infty\frac{U_n}{n!}b_i^{\+n}b_i^n.
\end{align}
Here the effective $N$-body interaction energy $U_N$ is a function of $\Omega$, $\Delta$, and $g_{\sigma\sigma'}$, which is determined so that the on-site energy from Eq.~(\ref{eq:effective}) matches the lowest eigenvalue of Eq.~(\ref{eq:on-site}) in each sector of up to $N$ bosons.%
\footnote{More specifically, the functional form of $U_N(\Omega,\Delta,g_{\sigma\sigma'})$ is implicitly obtained by equating $\sum_{n=1}^N\frac{N!}{n!(N-n)!}U_n$ with the lowest eigenvalue of the $(N+1)\times(N+1)$ matrix, $\<n|\H_i|n'\>$, where $|n\>\equiv\frac1{\sqrt{n!(N-n)!}}b_{\up i}^{\+n}b_{\down i}^{\+N-n}|\mathrm{vac}\>$ with $n=0,1,\dots,N$ are the basis states for $N$ bosons at site $i$.}

\begin{figure}[t]
\includegraphics[width=0.9\columnwidth,clip]{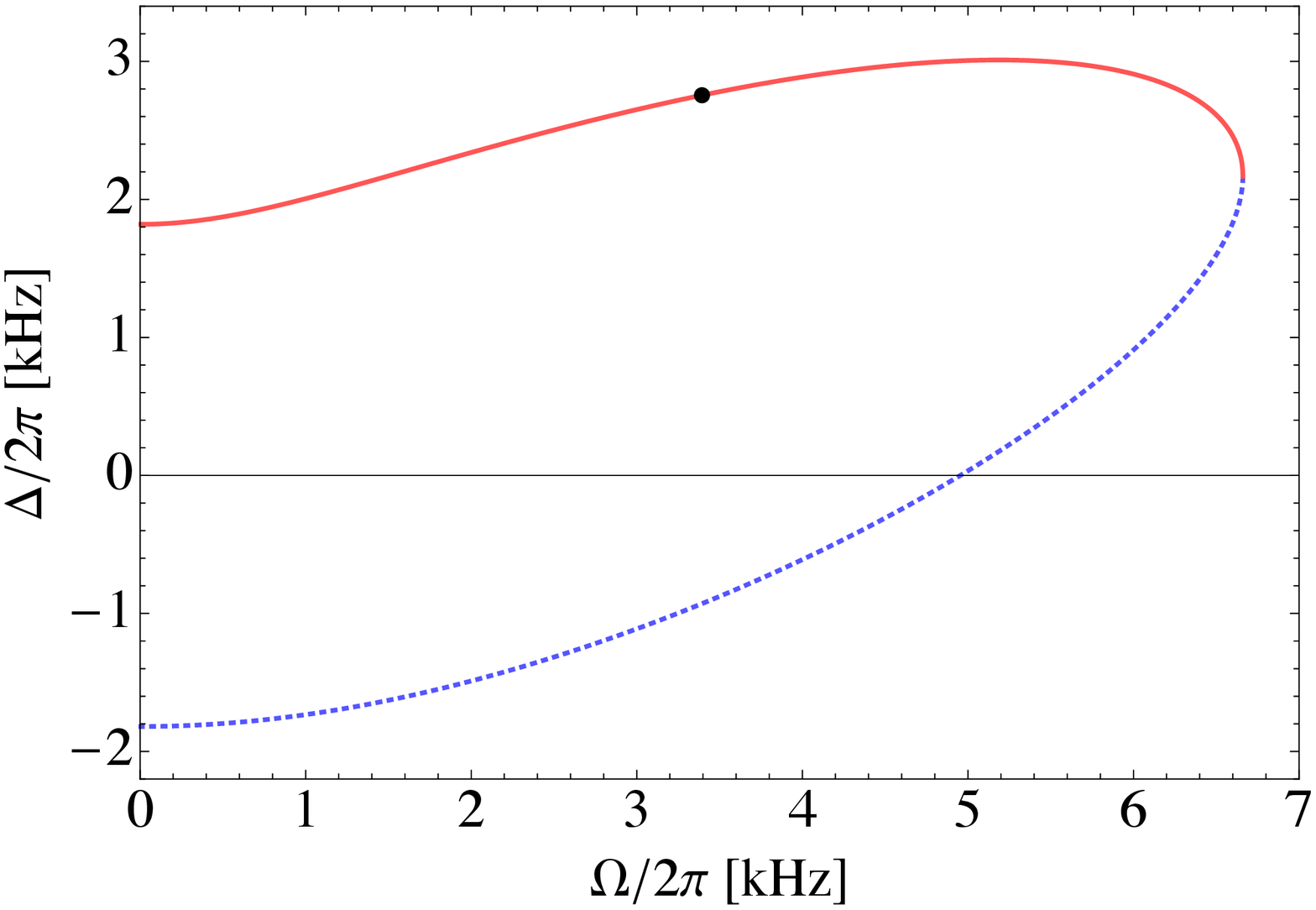}\bigskip\\
\includegraphics[width=0.94\columnwidth,clip]{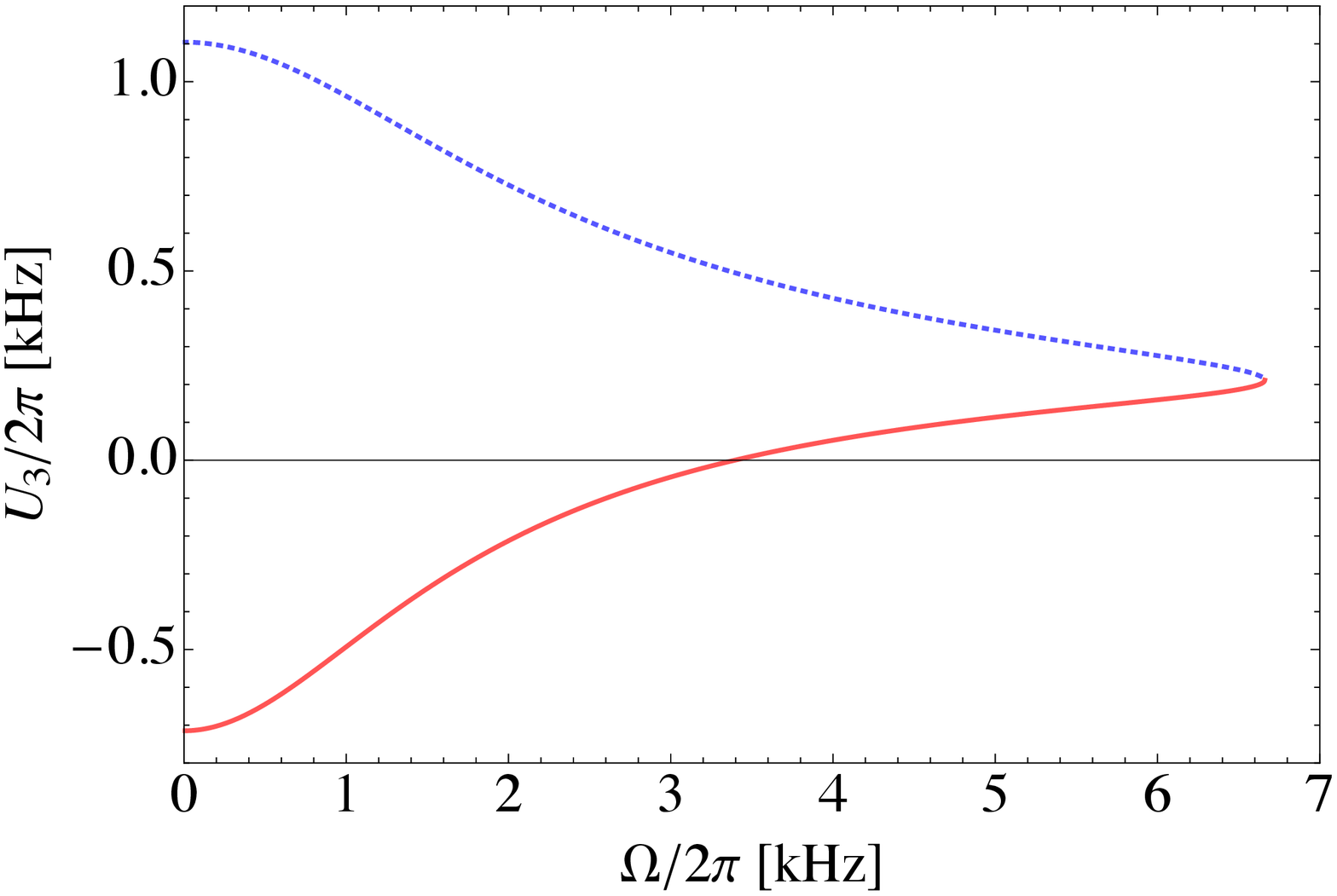}
\caption{\label{fig:rabi-detuning_U3}
(top panel) Curve in the plane of Rabi frequency $\Omega$ and detuning $\Delta$ along which the effective two-body interaction energy $U_2$ vanishes.
The dot marks the point at which both $U_2$ and $U_3$ vanish.
(bottom panel) Effective three-body interaction energy $U_3$ as a function of $\Omega$ with $\Delta$ simultaneously tuned to fix $U_2=0$.
The solid (dotted) curve corresponds to $\Delta$ along the solid (dotted) curve in the top panel.}
\end{figure}

Petrov in Ref.~\cite{Petrov:2014b} found that independent control of the effective two-body and three-body interaction energies is favorably achieved by choosing two spin components as $|F,m_F\>=|1,-1\>\equiv|{\up}\>$ and $|1,0\>\equiv|{\down}\>$ from $^{39}$K at $B\approx58$~G, where intra- and intercomponent scattering lengths read $a_{\up\up}\approx1.7$~nm, $a_{\down\down}\approx9.4$~nm, and $a_{\up\down}\approx-2.8$~nm~\cite{D'Errico:2007,Lysebo:2010}.
An optical lattice with the lattice constant $l\equiv\lambda/2=532$~nm and the anisotropic intensities $V_x=V_{y,z}/4=15\times2\pi^2/m\lambda^2$ in the axial ($x$) and radial ($y$ and $z$) directions leads to the intersite tunneling amplitude $t_x\approx2\pi\times30$~Hz with negligible $t_{y,z}$ as well as the on-site interaction energies $g_{\up\up}\approx2\pi\times1.1$~kHz, $g_{\down\down}\approx2\pi\times6.1$~kHz, and $g_{\up\down}\approx-2\pi\times1.8$~kHz in the harmonic approximation~\cite{Bloch:2008}.%
\footnote{Note the differences from Ref.~\cite{Petrov:2014b} by the factor of 2 because our oscillator lengths are $\ell_x=\sqrt2\ell_{y,z}\approx86$ nm.}
With this parameter set, $U_2$ in the plane of $(\Omega,\Delta)$ proves to vanish along the curve plotted in the top panel of Fig.~\ref{fig:rabi-detuning_U3} and its bottom panel then shows $U_3$ as a function of $\Omega$ with $\Delta$ simultaneously tuned to fix $U_2=0$.
Not only does $U_3$ vary in its magnitude, but also its sign changes at $(\Omega,\Delta)\approx2\pi\times(3.4,2.8)$~kHz from repulsive to attractive with decreasing $\Omega$.
This constitutes the realization of one-dimensional bosons without two-body but with tunable three-body interactions.
We note that, while effective four- and higher-body interactions also exist, they are irrelevant in our following discussion.

\section{\label{sec:few-body}Few-body bound states}
In the investigation of our system described by Eq.~(\ref{eq:effective}) with $U_2=0$ and tunable $U_3$, we first study the three-body problem.
While no difficulty arises to directly solve the lattice Schr\"odinger equation obeyed by the wave function $\Psi(i,j,k)$ with three coordinates of bosons, it is more instructive to express it as $\Psi(i,j,k)=e^{iPR}\psi_P(r_1,r_2)$ with new variables $R=(i+j+k)/3$, $r_1=\sqrt3(i-j)/2$, and $r_2=(i+j)/2-k$.
The lattice Schr\"odinger equation at zero center-of-mass momentum $P=0$ then reads
\begin{align}\label{eq:lattice}
E_3\psi_0(\r) = \left[-t_x\sum_{n=1}^3\Delta_n + U_3\delta_{\r,\0}\right]\psi_0(\r),
\end{align}
where $\Delta_n\psi_0(\r)\equiv\psi_0(\r+\e_n)+\psi_0(\r-\e_n)-2\psi_0(\r)$ is the discrete Laplacian with $\r=(r_1,r_2)$, $\e_1=(\sqrt3,1)/2$, $\e_2=(-\sqrt3,1)/2$, and $\e_3=(0,-1)$.
Interestingly, Eq.~(\ref{eq:lattice}) originating from three bosons in a one-dimensional lattice is equivalent to the Schr\"odinger equation for one particle in a triangular lattice with a potential energy concentrated at the origin,%
\footnote{Similarly, four bosons in a one-dimensional lattice proved to be equivalent to one particle in a body-centered-cubic lattice~\cite{Nishida:2010}.}
which reveals an intrinsic connection of our system with two-dimensional physics.
Consequently, the binding energy $E_3<0$ determined by
\begin{align}
\frac1{U_3} = \frac{\sqrt3}{16\pi^2}\int_0^{4\pi/\sqrt3}\!\!\int_0^{4\pi}\!d^2\p\,
\frac1{E_3-2t_x\sum_n[1-\cos(\p\cdot\e_n)]}
\end{align}
has a solution for an arbitrary three-body attraction $U_3<0$.
In particular, we find the exponentially small binding energy
\begin{align}\label{eq:binding}
E_3 \to -72\,t_x\exp\!\left(-4\sqrt3\pi\frac{t_x}{|U_3|}\right)
\end{align}
in the weak attraction limit $U_3/t_x\to-0$.

When the three-body attraction is so weak that the bound state extends over many lattice sites, low-energy properties of the system are universal and can be described by the theory in the continuum limit.
By making the replacement of $t_xl^2\to1/2m$, $b_i/\sqrt{l}\to\phi(x)$, and $U_nl^2\to u_n/m$, the continuum limit reduces the lattice Hamiltonian in Eq.~(\ref{eq:effective}) with $U_2=0$ to
\begin{align}\label{eq:continuum}
H_\mathrm{cont} = -\int\!dx\,\phi^\+(x)\frac{\d^2}{2m}\phi(x)
+ \frac{u_3}{6m}\int\!dx\,[\phi^\+(x)]^3[\phi(x)]^3.
\end{align}
Here four- and higher-body interaction terms all disappear because they are accompanied by positive powers of the lattice constant $l\to0$, i.e., irrelevant in the sense of the renormalization group.
While the three-body coupling constant $u_3<0$ is dimensionless, this continuum theory admits bound states by virtue of the dimensional transmutation~\cite{Coleman:1973}.
The emergent scale should be matched with the binding energy of three bosons from the microscopic lattice model.

For the sake of generality, we study the $N$-body problem described by the following Schr\"odinger equation derived from Eq.~(\ref{eq:continuum}):
\begin{align}
E_N\Psi(\x) = \left[-\sum_{n=1}^N\frac{\d_n^2}{2m}
+ \sum_{1\leq i<j<k\leq N}\frac{u_3}{m}\,\delta(x_{ij})\delta(x_{jk})\right]\Psi(\x),
\end{align}
where $\x=(x_1,\dots,x_N)$ is a set of $N$ coordinates of bosons and $x_{ij}=x_i-x_j$ is the interparticle separation.
The corresponding momentum-space integral equation for $E_N=-\kappa_N^2/m<0$ in the center-of-mass frame reads
\begin{subequations}\label{eq:integral}
\begin{widetext}\begin{align}
& \left[\frac1{u_3} + \iint_{-\Lambda}^\Lambda\!\frac{dp_2dp_3}{(2\pi)^2}
\frac1{\kappa_N^2+\frac12\bigl(\sum_{n=2}^Np_n\bigr)^2+\sum_{n=2}^N\frac{p_n^2}{2}}\right]
\tilde\psi_0(\p\text\textbackslash\{p_1,p_2,p_3\}) \notag\\
&= -\iint_{-\infty}^\infty\!\frac{dp_2dp_3}{(2\pi)^2}
\frac1{\kappa_N^2+\frac12\bigl(\sum_{n=2}^Np_n\bigr)^2+\sum_{n=2}^N\frac{p_n^2}{2}}
\sum_{1\leq i<j<k\leq N}^{(i,j,k)\neq(1,2,3)}
\tilde\psi_0(\p\text\textbackslash\{p_i,p_j,p_k\})\bigg|_{p_1\to-\sum_{n=2}^Np_n},
\end{align}\end{widetext}
where $\tilde\psi_P(\p\text\textbackslash\{p_i,p_j,p_k\})$, with its argument referring to $\p=(p_1,\dots,p_N)$ with $\{p_i,p_j,p_k\}$ excluded, is the Fourier transform of $\Psi(\x)|_{x_i=x_j=x_k\equiv X}$ with the momentum conjugate to $X$ being $P-\sum_{n\neq i,j,k}p_n$ for the center-of-mass momentum to be $P$.%
\footnote{Integral equations of the same type were derived for mass-imbalanced $N+1$ fermions in general dimensions~\cite{Pricoupenko:2011,Bazak:2017} as well as for $N$ bosons in two dimensions~\cite{Bazak}.}
The integral in the square brackets on the left-hand side is logarithmically divergent and thus is cut off by $\Lambda$.
This cutoff dependence can be eliminated by defining the three-body coupling constant as $u_3=-\sqrt3\pi/\ln(a_3\Lambda)$ so that we obtain
\begin{align}\label{eq:coupling}
\left[\frac1{u_3} + \cdots\right] = -\frac1{\sqrt3\pi}
\ln\!\left[a_3\sqrt{\kappa_N^2+\frac16\left(\sum_{n=4}^Np_n\right)^2+\sum_{n=4}^N\frac{p_n^2}{2}}\right]
\end{align}
\end{subequations}
in the limit of $\Lambda\to\infty$.

Here $a_3$ is referred to as a three-body scattering length.
Its physical meaning can be revealed by solving Eq.~(\ref{eq:integral}) for $N=3$, which reduces to $\ln(a_3\kappa_3)=0$, and thus the binding energy of three bosons is provided by $E_3=-1/ma_3^2$.
By matching it with Eq.~(\ref{eq:binding}) from the microscopic lattice model, we find that the three-body scattering length expressed in terms of the microscopic lattice parameters is
\begin{align}
a_3 = \frac{l}{6}\exp\!\left(2\sqrt3\pi\frac{t_x}{|U_3|}\right) \gg l
\end{align}
in the weak attraction limit $U_3/t_x\to-0$.
Because the three-body scattering length is the only dimensionful parameter in Eq.~(\ref{eq:integral}), the binding energies of $N\geq4$ bosons, if they exist, must also be in the form of $E_N=-\#/ma_3^2$.
Consequently, their ratios to $E_3$ are universal, i.e., independent of the microscopic lattice parameters.

In order to determine these universal numbers, the $(N-3)$-dimensional integral equation in Eq.~(\ref{eq:integral}) needs to be solved numerically.
In particular, we find three bound states of $N=4$ bosons with their binding energies provided by
\begin{align}
\ln(\kappa_4/\kappa_3) \approx 6.77246,\quad 2.46114,\quad 0.376644.
\end{align}
While we will not attempt to solve Eq.~(\ref{eq:integral}) for $N\geq5$, we now show with an alternative approach that $N\gg1$ bosons form a many-body bound state (droplet) and its ground-state energy grows exponentially with increasing particle number.

\section{\label{sec:many-body}Many-body droplet}
Toward this end, we employ the approach developed by Hammer and Son for self-bound attractive bosons in two dimensions, which relies on the classical field theory applicable to a large number of bosons~\cite{Hammer:2004,Bazak}.
In fact, the Gross-Pitaevskii equation with the right choice of coupling was proven to provide the ground-state energy and number density exactly for trapped repulsive bosons both in three dimensions~\cite{Lieb:1998,Lieb:2000} and in two dimensions~\cite{Lieb:2001a,Lieb:2001b}.
Here we proceed by considering the same to be true for self-bound attractive bosons in one dimension, which can be confirmed explicitly in the case of two-body attraction~\cite{Castin:2001}.%
\footnote{For one-dimensional bosons with a two-body attraction $u_2<0$, the minimization of
$E_N = -\int\!dx\,\phi^\+(x)\frac{\d^2}{2m}\phi(x)
+ \frac{u_2}{2m}\int\!dx\,[\phi^\+(x)]^2[\phi(x)]^2$
with respect to $\phi(x)$ under $\int_{-\infty}^\infty\!dx\,|\phi(x)|^2=N$ leads to $E_N=-\frac{u_2^2N^3}{24m}$ and $|\phi(x)|^2=\frac{N}{2R}\frac1{\cosh^2[(x-x_0)/R]}$ with $R=\frac2{|u_2|N}$, which are consistent with the exact results for $N\gg1$~\cite{Castin:2001}.}

The ground-state energy of $N\gg1$ bosons is determined so as to minimize the energy functional in Eq.~(\ref{eq:continuum}) with respect to $\phi(x)$ regarded as a classical field satisfying $\int_{-\infty}^\infty\!dx\,|\phi(x)|^2=N$.
This normalization condition is conveniently incorporated by expressing the wave function as
\begin{align}
\phi(x) = \sqrt{\frac{N}{CR}}\,\varphi\!\left(\frac{x}{R}\right),
\end{align}
where $C\equiv\int_{-\infty}^\infty\!d\xi\,[\varphi(\xi)]^2$.
The dimensionless real function $\varphi(x/R)$ must vanish at $|x|\gg R$ for the convergent normalization integral and thus $R$ sets the size of $N$-boson droplet.
In order to minimize the energy with respect to $R$, it is essential to take into account the logarithmic scale dependence of coupling through the renormalization~\cite{Lieb:2001a,Lieb:2001b,Hammer:2004}.
This can be observed in Eq.~(\ref{eq:coupling}) where the bare coupling constant on the left-hand side is renormalized by quantum mechanical corrections to turn into the scale-dependent running coupling on the right-hand side.
Because characteristic momentum of bosons confined in a droplet of size $R$ is $p_n\sim R^{-1}$ and also $\kappa_N\sim R^{-1}$ as confirmed later, the three-body coupling constant should be replaced as $u_3\to-\sqrt3\pi/\ln(a_3/R)$ to the leading logarithmic accuracy.
Consequently, the energy functional to be minimized with respect to $R$ and $\varphi(\xi)$ reads
\begin{align}\label{eq:functional}
mE_N = \frac{A}{2C}\frac{N}{R^2}
- \frac{\sqrt3\pi}{6\ln(a_3/R)}\frac{B}{C^3}\frac{N^3}{R^2},
\end{align}
where $A\equiv\int_{-\infty}^\infty\!d\xi\,[\varphi'(\xi)]^2$ and $B\equiv\int_{-\infty}^\infty\!d\xi\,[\varphi(\xi)]^6$.

The three-body attraction in Eq.~(\ref{eq:functional}) tends to shrink the droplet toward $R\to0$ but simultaneously the three-body running coupling decreases.
Eventually, when $\ln(a_3/R)\gtrsim N^2$ is reached, the kinetic energy dominates over the three-body attraction and thus stabilizes the droplet from collapsing.
The optimal size of droplet proves to be
\begin{align}\label{eq:size}
R = a_3\exp\!\left(-\frac\pi{\sqrt3}\frac{B}{AC^2}N^2 + \cdots\right)
\end{align}
and minimizes the energy at $mE_N\propto-1/NR^2$.
This energy should be minimized further with respect to $\varphi(\xi)$, i.e., the shape of droplet, which is achieved by maximizing the ratio $B/AC^2$ in the exponent of Eq.~(\ref{eq:size}).
By introducing the rescaled $\varphi(\xi)=(2B/C)^{1/4}\tilde\varphi(\tilde\xi)$ with $\tilde\xi=(8A/C)^{1/2}\xi$, the extremization of $B/AC^2$ with respect to $\varphi(\xi)$ leads to
\begin{align}
4\tilde\varphi''(\tilde\xi) + 3[\tilde\varphi(\tilde\xi)]^5 - \tilde\varphi(\tilde\xi) = 0,
\end{align}
which with the boundary condition $\tilde\varphi(\tilde\xi\to\pm\infty)\to0$ is solved by $\tilde\varphi(\tilde\xi)=1/\sqrt{\cosh(\tilde\xi-\tilde\xi_0)}$, where $\tilde\xi_0$ is a constant of integration.
Consequently, we find $B/AC^2=4/\pi^2$ at its maximum and the ground-state energy of $N$-boson droplet is provided by
\begin{align}\label{eq:energy}
E_N = E_3\exp\!\left(\frac8{\sqrt3\pi}N^2 + \cdots\right)
\end{align}
and the number density by
\begin{align}
|\phi(x)|^2 = \frac{N}{\pi R\cosh[(x-x_0)/R]}.
\end{align}
These expressions, where potential $O(N)$ corrections are expected in the exponents of Eqs.~(\ref{eq:size}) and (\ref{eq:energy}), are valid for $N\gg1$ as long as the droplet is so dilute that the mean interparticle separation is much larger than the lattice constant.
This requires $R/N\gg l$, that is,
\begin{align}
1 \ll N \ll \sqrt{\frac{3\pi^2}{2}\frac{t_x}{|U_3|}}
\end{align}
in terms of the microscopic lattice parameters.

\section{Conclusion}
In this paper, we presented comprehensive studies of one-dimensional bosons with a weak three-body attraction, from their lattice realization (Sec.~\ref{sec:lattice}) through few-body bound states (Sec.~\ref{sec:few-body}) to a many-body droplet stabilized by the quantum mechanical effect (Sec.~\ref{sec:many-body}).
In particular, we showed that their binding energies relative to that of three bosons are all universal and found that the ground-state energy of the dilute droplet grows exponentially as in Eq.~(\ref{eq:energy}) with increasing particle number.
Our predictions are in principle testable in ultracold-atom experiments with coupled two-component bosons in an optical lattice.
We expect further unique phenomena to be revealed in novel systems without two-body but with tunable three-body interactions, which can be an important research frontier in ultracold atoms.

\begin{table}[b]
\caption{\label{tab:droplets}
Ground-state energies of universal $N$-boson droplets for $N\gg1$ with few-body attractions (columns) in various dimensions (rows).
The (semisuper) Efimov effect indicates the universality not in the ground state but only in higher excited states.
The other systems marked by dashes are considered to be nonuniversal.}\smallskip
\begin{ruledtabular}
\begin{tabular}{cccc}
Dim.\textbackslash Att. & Two-body & Three-body & Four-body \\[2pt]\hline\\[-10pt]
1D & $N^3$~\cite{McGuire:1964} & $e^{8N^2/\sqrt3\pi}$~[present work] & Efimov~\cite{Nishida:2010} \\
2D & $e^{2.148N}$~\cite{Hammer:2004} & semisuper Efimov~\cite{Nishida:2017} & --- \\
3D & Efimov~\cite{Efimov:1970} & --- & ---
\end{tabular}
\end{ruledtabular}
\end{table}

Finally, our finding herein advances our perspective on the fates of attractive bosons with few-body interactions in various dimensions.
Two-body attractions in one and two dimensions as well as our three-body attraction in one dimension lead to quantum droplets of bosons stabilized by balancing kinetic and potential energies dominant at short and long distances~\cite{McGuire:1964,Hammer:2004}.
These quantum droplets, as long as they are dilute, exhibit universal properties including their ground-state energies summarized in Table~\ref{tab:droplets}.
On the other hand, bosons with two-, three-, and four-body attractions in three, two, and one dimensions, respectively, suffer from the Efimov or semisuper Efimov effect~\cite{Efimov:1970,Nishida:2017,Nishida:2010}.
This indicates that the ground state is nonuniversal already in few-body sectors but instead the universal scaling law emerges in binding energies of higher excited states.
It is both interesting and instructive to observe in Table~\ref{tab:droplets} that the Efimov effects lie right at the boundary dividing the universal (upper left) and nonuniversal (lower right) systems, which may help us to develop deeper insights into the universality in quantum few-body and many-body physics.

\acknowledgments
This work was supported by JSPS KAKENHI Grants No.~JP15K17727 and No.~JP15H05855.

\end{document}